# PULSED DEMO DESIGN ASSESSMENT STUDIES

T N Todd

EURATOM/CCFE Fusion Association, Culham Science Centre, Abingdon, OX14 3DB, UK
tom.todd@ccfe.ac.uk

*Now that ITER is under construction, interest is increasing in the specification and design of the successor machine, a Demonstration Power Plant (DEMO), which in Europe is coordinated by the EFDA Power Plant Physics and Technology programme. This paper summarises the work carried out for EFDA in 2011-2012 on design issues pertinent to a pulsed version of DEMO, intended to be implemented with little or no extrapolation of technology available today. The work was carried out by the Euratom Fusion Associations CCFE, CEA, CRPP, ENEA and KIT, and in addition to a review of recent relevant literature addressed systems code analyses (pulse length vs. size), erosion of plasma facing components, thermomechanical fatigue in the blanket and first wall, a range of energy storage issues, and fatigue life improvements in $Nb_3Sn$ CICC superconductors.*

## I. BACKGROUND

In 2011, the EFDA Power Plant Physics and Technology group launched a Call for Participation in DEMO Design Assessment Studies, including consideration of a pulsed version of DEMO. Five of the Euratom Fusion Associations subsequently undertook this work, reporting it in January 2012. The aim was not to produce an outline design of a pulsed DEMO reactor but to identify some of the advantages and disadvantages of a pulsed device compared to the characteristics of a steady state reactor, still the ultimate aim of the EFDA PPPT programme. Much was learned from the analyses of the reference pulsed reactor and also from some explorations of variants offering reduced demands on technology, and as a result some of these studies are being carried forward by EFDA PPPT within its work-plan for 2012-2013. The many important features common to both pulsed and steady state tokamak reactors, such as tritium breeding, remote handling, safety, waste management and economic issues were not within the scope of these contracts. This paper summarises the key findings of the 2011-2012 studies and is structured to reflect the individual contributions of the different Associations. Inevitably it cannot cover all the aspects addressed in each topic but contact can be facilitated with EFDA PPPT for those readers wishing to know more of the full scope or greater detail of the work reported here.

## II. LITERATURE SURVEY (ENEA Frascati)

This survey concentrated on publications made since 2007 of relevance to pulsed DEMO. One driving force for a pulsed DEMO is the large recirculating power necessary to support the current in a steady-state option. Even with an optimistic current drive efficiency (meaning $I_P R_0 n_e / P_{CD}$, $n_e$ being the line averaged density) of $0.7 \times 10^{20}$ A/Wm$^2$ (a more conventional value being $0.4 \times 10^{20}$ A/Wm$^2$) and a bootstrap fraction of ~75%, over 100MW of heating power is required, which might well demand ~250MW or more of electrical input power.[1,2] Obviating this problem and its heating system development implications is one clear advantage of a pulsed DEMO, leading to the ongoing re-evaluations of the pulsed option. A consistent conclusion is that to accommodate a sufficient solenoid diameter (i.e. sufficient poloidal flux swing) to permit a pulse length of some hours requires a somewhat larger reactor than the steady state versions, where some form of plasma current drive is assumed to assist start-up and sustain the flat-top.

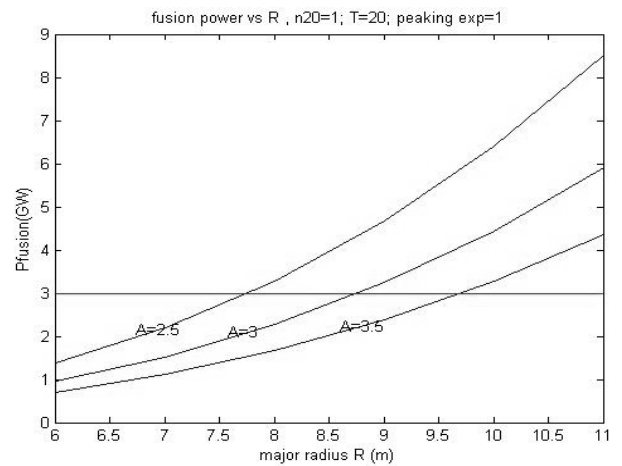

Fig. 1. The variation of fusion power with major radius and aspect ratio ($B_0$=8.0T, A=3.5, $n_e$=0.8$n_{Gr}$ )

Typically a large major radius of around 9m is recommended, both in the literature[3] and in a simple reactor model produced as part of this work, as shown in Figure 1. It appears to be possible, if desired, to reduce the fusion power output to around 1GW in the pulsed version compared to the canonical 3GW of the steady-state one, still achieving ignition (defined as a very high Q to keep the steady state heating power low) with ITER98(Y)2 confinement scaling.

Several caveats arose, including the uncertainty of reliably maintaining both an Internal Transport Barrier and sufficient alpha confinement with hollow current density profiles, avoiding unfavourable self-organisation effects and demonstrating satisfactory burn control techniques, including practicable diagnostics and control actuator systems. It was also noted that while tokamaks have made good progress in achieving many of the dimensionless parameters required of a reactor, such as confinement $H_H$, and pressure $\beta_N$, their simultaneous demonstration necessarily awaits a machine with reactor-scale parameters including normalized ion Larmor radius $\rho^*$ and collisionality $\nu^*$, i.e. a reactor.[4]

### III. FIRST WALL HEATING AND EROSION (KIT)

It is universally accepted that disruptions must be vanishingly rare in DEMO, but it is unlikely that the regulators could be convinced that they will never occur and so methods must be both identified and proven that allow their consequences to be mitigated effectively. The transients associated with ramping up the current, moving from limiter to divertor mode, heating to obtain significant alpha power in a stable scenario, followed after the burn by a run-down of heating power, pressure, density and current, are generally considered to be more likely to lead to disruptions than continuous operation in a steady state reactor. Thermo-mechanical fatigue of the plasma facing components will also be higher with the pulsed option. If steady state involved smaller stability margins against plasma instabilities, this might bring the predicted disruptivity of the two design classes closer together, but usually it is predicted that pulsed reactors would be the more disruptive. In this study, the thermal loads associated with three types of disruptive events were considered: (a) a "hot" Vertical Displacement Event in which the hot plasma moves rapidly until it strikes the plasma facing components in the vessel, producing 50-100MJ/m$^2$ in ~1s; (b) a "cold" Vertical Displacement Event in which the vertical displacement occurs after a disruption has cooled the plasma, producing 30-50MJ/m$^2$ in 0.3-1s; and (c) the generation of runaway electrons during a disruption of any type, producing ~50 MJ/m$^2$ in 0,05-0.3s. Vapour screening effects were included in the analysis. Electromagnetic loads (vital for designing the details of castellations), segmentation, wall fixings, engineered current paths etc necessary to cope with disruption-induced currents, were not within the required scope of this study.

In the 'cold' VDE case, the analyses show that increasing the thickness of the W armour somewhat can keep the maximum heat flux and temperature below the critical levels for both the W cladding and the EUROFER substrate. However for the 'hot' VDE case, the energy deposition into the W armour is very shallow (~ nm) and causes surface melting and evaporation.

Runaway electrons can in principle be prevented by massive gas injection, but the quantity of gas required may be very large. If mitigation is unsuccessful, analysis suggests that runaway electrons could deposit 40 MJ in 0.05-1s over a very small area, ~0.8 m$^2$. An analysis of the effect of this on a EUROFER blanket with tungsten armour concludes that the minimum thickness of armour required to prevent the EUROFER from excessive creep (which will occur at ~1000K for unirradiated material) or thermal stress is ~1.4cm, as shown in Figure 2. However, such a thick layer of armour is likely to suffer surface melting, sufficiently to eject droplets into the plasma. Clearly, hot VDEs and runaway electrons require reliable mitigation strategies to ensure first wall protection.

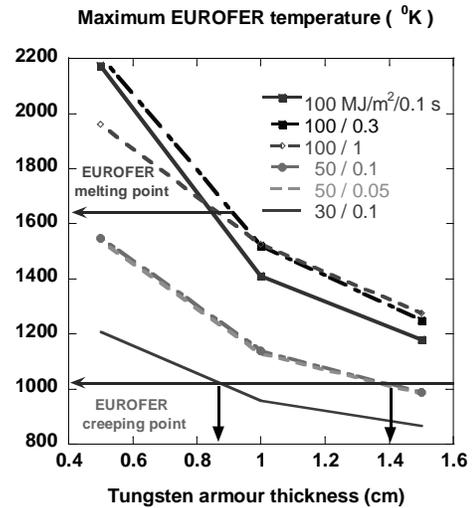

Fig. 2. The effect of tungsten armour thickness on the peak temperature suffered by the EUROFER in disruption transients.

The erosion of the first wall is mainly due to sputtering by fast neutrals created by charge exchange, while the erosion rate of divertor plates is essentially determined by sputtering due to fuel and impurity ions. Estimations of sputtering that assume normal incidence, failing to take into account the factor of ~10 increase of sputtering yield at shallow angles of incidence, will considerably underestimate the rate of erosion.[5]

Preliminary calculations suggest total sputtering erosion by charge-exchange neutrals of W armour on the first wall could reach ~1mm during one full power year, for a particle flux of fuel species and helium ash (and no significant content of high Z impurities) of $10^{23}/m^2$s with edge temperatures of ≥200eV. Since sputtering from the divertor target is dominated by ions (unless the divertor plasma is fully detached), it will be exacerbated by the ion acceleration given by the product of sheath potential and ion charge state, potentially leading to about 10mm/year erosion rate of the tungsten armour if a significant impurity content and high electron temperature at the sheath cannot be avoided. However on-going divertor experiments and ITER developments are expected to mitigate these parameters, while future analysis could take account of the effect of the sheath acceleration normal to the surface in ameliorating the strong angular dependence of the sputtering yield elaborated above.

**IV. SYSTEMS CODE STUDIES (CEA)**

In this analysis, CEA sought a set of reactor design parameters very different from those provided as a reference set for the other work reported in this paper but consistent with their proposal that an early DEMO might have only 1 GW of fusion power, with only 4.8T central toroidal field, $\beta_N$~1.7 and no current drive at all. The validity of the CEA systems code HELIOS used in this study was demonstrated by benchmarking it against the (PROCESS) reference parameter set, as shown in Table I.

In the proposed variant, a very low residual nuclear heating of the superconductor was assumed, ~ 50W/m$^3$, consistent with a 1.9m shielding thickness, given appropriate selection of the shield materials. Together with moderate choices for the maximum TF on the coil of 12T and a cable current density of 50A/mm$^2$, this allows a temperature margin of 2K, much larger than the ~0.7K of the ITER design. The peak field in the central solenoid was left at the ITER value of 13T.

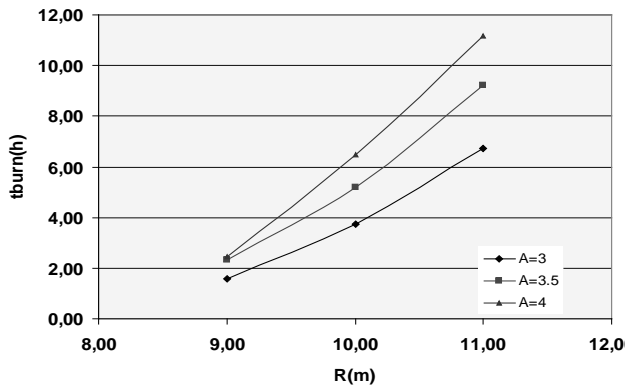

Fig. 3. The increase of pulse length with major radius and aspect ratio.

Since no suitably high efficiency current drive technology or proven superconducting cable designs are available at present to build a more "standard" 3 GW, 7.45 T reactor, the CEA proposal can be viewed as a reasonable near-term demonstration machine. The cost of electricity would be high due to the capital cost of a machine of modest output power whose major radius is ~9 m. The recirculating power, however, would be small because of the absence of current drive, and Q would seem likely to reach ~80 although for the assumptions and parameter combinations considered, Q falls as R is increased. A slight improvement to pulse length would be obtained if the plasma was heated during the start-up phase, to reduce the resistivity and therefore the flux consumption during ramp-up. The burn time would then be ~3 hours, depending on the major radius as shown in Figure 3, which demonstrates a trend similar to that of previous studies on pulsed tokamak fusion reactors such as those noted in Section II above where a much higher nominal toroidal field was assumed (8.0T instead of 4.8T), which reduces the required size for a given fusion power since $P_{fusion}$ is roughly proportional to the product of volume and $B^4$ for a given aspect ratio and β.

TABLE I. Reference benchmark analysis

|  | PROCESS (ref) | HELIOS |
|---|---|---|
| R (m) | 9.6 | 9.6 |
| $B_t$ (T) | 7.45 | 7.45 |
| $\beta_N$ | 2.6 | 2.6 |
| $\kappa_{95}$ | 1.7 | 1.7 |
| A | 4 | 4 |
| $n_e/n_G$ | 1.0 | 1.0 |
| $q_{95}$ | 3 | 3 |
| $T_i/T_e$ | 1.0 | 1.0 |
| $P_{fus}$ (MW) | 2700 | 2705 |
| $I_p$ (MA) | 18 | 18.1 |
| $T_e$ (keV) | 19.0 | 19.17 |
| $n_e$ ($10^{19}/m^3$) | 10 | 9.68 |
| Q |  | 1000 |
| H | 1.2 | 1.2 |
| $Z_{eff}$ | 1.95 | 1.95 |

**V. SUPPORT STRUCTURE FATIGUE (CCFE)**

The aim of this task was to explore the cost implications of trading off a reduction in the size and cost of the central solenoid and associated power supply of a pulsed DEMO against the resulting shortened pulse length, i.e. the increase in the number of lifetime pulses and corresponding strengthening of the load assembly components to provide an adequate fatigue life. The first approach tried was to assume that the usual engineering constraints of the PROCESS systems code[1] were to suit a fatigue life of around 100 pulses (a few maintenance outages per year over a 30 year life), which for the pulsed machine would have to become ~37,000 pulses. It transpired that the slope of the S-N curves for typical structural materials implied a need for a greater thickness of coil support structures to accommodate the increased fatigue life than could be made consistent with the overall design tenets of PROCESS, if simple thickening was used to reduce the stress. A stylised radial build consistent with the tokamak reactor parameters and assumptions used in PROCESS is shown in outline in Figure 4, where it will be apparent that the thickness of the magnet and (integral) structural components (arrowed) could not be raised by a factor of greater than ~2 before the gaps between nested components closed. S-N data was collected on a range of exotic alloys to see if a different cost-benefit paradigm could be inferred by increasing the stress capability at 37,000 cycles simply by increasing the cost of the alloy, but the costs proved both hard to quantify and where they could be determined, prohibitively expensive.

Accordingly the assumptions about the basis of the radial build implicit within PROCESS were revisited, noting that this code does not attempt to incorporate design details or any S-N characteristics of the materials assumed. Comparative runs were made now with the assumption that on the basis of the key reactor parameters given to it, the radial build produced by PROCESS could be taken to suit a life of 37,000 pulses, allowing the desired cost-benefit variation to be explored. Noting that since PROCESS does not embody any actual design for the coil support structure, this is considered reasonable because the choice of alloy, stress-reduction features etc remains open for optimization in any real design.

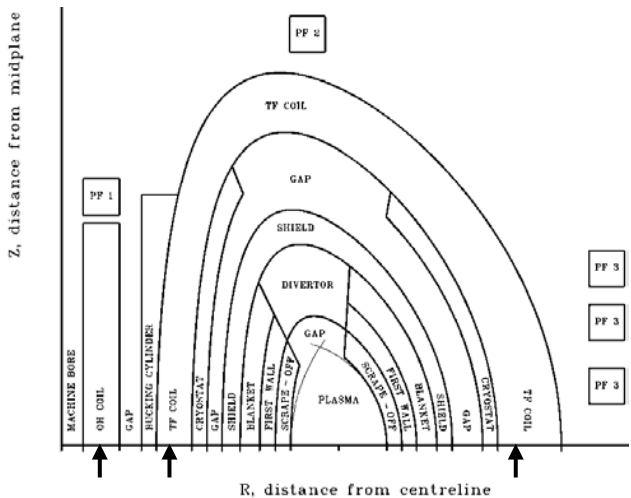

Fig. 4. Schematic guide to the interpretation of PROCESS radial build parameters (not to scale). The arrows identify the radial zones corresponding to the windings and support structures for the central solenoid and toroidal field coils.

TABLE II. Comparison of reference and enlarged-solenoid reactor designs explored by PROCESS

| Parameter | Reference | Enlarged solenoid |
|---|---|---|
| Available flux swing (Wb) | 873 | 1058 |
| Pulse length (s) | 12890 | 19040 |
| Major Radius (m) | 9.58 | 9.88 |
| Minor Radius (m) | 2.395 | 2.41 |
| Aspect Ratio | 4.0 | 4.1 |
| TF on axis (T) | 7.45 | 7.45 |
| Plasma Current (MA) | 18.0 | 17.6 |
| Average Temperature (keV) | 19 | 19 |
| Average Density ($10^{20}$m$^{-3}$) | 1.0 | 1.0 |
| Confinement time (s) | 3.94 | 3.90 |
| Net heating power MW) | 412 | 434 |
| Fusion Power (GW) | 2.70 | 2.82 |
| Thermal Power (GW) | 3.25 | 3.39 |
| Net Electric Power (GW) | 1.03 | 1.08 |
| COE (m$/kWh) | 151 | 149 |

After some iteration to recover much the same output power, a design was arrived at with the solenoid bore area increased by a factor of 1.56, producing the same net electrical output of ~1.05GW but with the pulse duration increased from 13,000s to 19,000s, as shown in Table II. Surprisingly, to within 1.3% (which is well below the uncertainties of the analysis), the predicted cost per kWh of the net electricity production was the same. The increased cost of the solenoid counteracted the reduction in thickness of the structural components, i.e. the design point chosen is either near a minimum or the variation in net costs for this trade-off is weak.

## VI. ENERGY STORAGE (CCFE)

When a pulsed reactor is between fusion burns (i.e. in the "dwell" period), the electricity grid will need to compensate in some way, either using energy storage, or by rapidly ramping up spare capacity such as gas turbines or wind turbines held in reserve. In addition, power is required to recharge the OH solenoid and to restart the fusion reactor The power required for starting up the reactor peaks at ~0.5GWe if the dwell time is 500s, based on analyses from the system code "PROCESS" and a simple POPCON (reactor Power OutPut CONtours) code. Longer dwell times allow a slower recharge of the OH solenoid and slower ramp-up, requiring less power. There will also be utility grid constraints ("rules") that must be observed. The operational rules or constraints used by the UK National Grid in 2010 include:

- Max. infrequent Infeed Loss (sudden): 1320 MW
- Max. capacity change rate: 50MW/min
  (120MW/min in Europe)

Accordingly, an energy storage system could be specified in three possible ways:

1. To compensate the missing electric power output to the grid during the dwell time, and to supply power for pulse initiation. This does not need to be on-site.
2. To supply power for pulse initiation only, arguably on-site.
3. To stretch out the ramp-up and ramp-down periods at the beginning and end of each pulse to suit grid rules, most appropriately on-site.

The grid and its operational constraints are expected to look very different by the time DEMO is brought on line, since energy storage will be required to make use of renewable energy sources predicted to be abundant by then.

At the end of a pulse, the fusion power is likely to run down in 10-100 seconds. There will be a degree of thermal storage intrinsic to the steam plant, and residual heat created in the blanket due to radioactive decay, but this will not be enough to maintain the steam circuit at a temperature high enough for useful electricity generation for more than a few minutes. Full electricity output will therefore stop within ~100s, giving a ramp-down rate of at least 600 MW/min, which would not be acceptable in any present grid for routine operation. Enough energy storage to stretch the ramp-down to eight minutes would be needed to satisfy even the European capacity change rate limit applied today, and this would have to be on-site.

A similar storage capacity would be required to smooth the power ramp-up transient. In addition, pulse initiation will require several hundred MW, ramped-up in ~100s. A pulsed reactor might perhaps have a dwell time of about 15 minutes, so if the maximum permissible capacity change rate remains the same as today, the energy storage required for it is of the same order of magnitude as the total power of the reactor – 1 GWe - for ~15 minutes, and this needs to be on-site in order to satisfy the grid requirements. It is a moot point that such an energy storage system should be able to earn income by providing a valuable load-leveling function for the grid during the many hours per reactor pulse that it was otherwise standing idle, but that would have to be balanced against the penalties of not having it available immediately if/when needed for its original purpose.

World-wide, many energy storage systems have been installed on a MW scale, but only two types are in use on a GW scale: pumped water and compressed air. More than 200 pumped hydroelectric storage systems in use all over the world provide a total of more than 100 GW of generation capacity. The only other technology currently available on a very large scale is Compressed Air Energy Storage (CAES). There are two CAES units in the world, the larger of which can generate 290 MW for 2 hours, but several more are planned or under construction, including the Norton (Ohio) project,[6] intended to have 2.7GW, 43 GWh capability. Both of these technologies, however, require specific geological features which are unlikely to be available on-site, but they are suitable for maintaining the overall power required by the grid.

Molten salt storage is probably the third most advanced technology in the field. The Valle 1 and Valle 2 plants of the Andasol[7] project in Spain are now complete – together equipped with a storage capacity of 100 MW for 7 hours. Capital costs are reckoned to be $50 to $100 per kWh. The cost for the full requirement for a fusion plant (option 1 above, taken as 250MWhr) would then be $12.5-25 million, a surprisingly low figure. The application of such a figure, if substantiated, would have to be based on the assumption that the very much higher power (than Andasol's 100MW) steam turbines and generators needed for smoothing the fusion reactor output would be shared between the main power station and the thermal storage system.

Superconducting magnet electrical storage (inductive energy storage) is also under development but seems unlikely to attain the necessary energy capacity, except perhaps for the less onerous task of smoothing high speed transients associated with the plasma heating and vertical stability control systems. If suitably developed, it would at least be able to make use of the cryoplant needed for the tokamak coils. There are also developmental systems featuring heat engines to move heat energy in and out of a "hot gravel" silo in one case, and a liquid air reservoir in another, and large-scale energy storage is in general a rapidly developing field driven by the interest in renewables with their intermittent power generation.

## VII. USE OF START-UP PLASMA HEATING SYSTEM FOR FLAT-TOP CURRENT DRIVE

POPCON and PROCESS code runs were used to find a trajectory in density and temperature that requires the minimum heating power necessary to ramp the plasma through the initial L-mode regime into H-mode and on up to ignition or an operating point with very high power gain, Q. It was noted that whatever plausible path is used in the ramp-up, the maximum heating power required is, for a steady-state reactor, at its chosen operating point, owing to the large power required for current drive. The exact shape of the POPCON plot and the scaling of the L-H threshold power are therefore not critical if such an operating point was the aim. On the other hand, a pulsed reactor with no current drive can be operated with little or no additional heating by choosing an operating point closer to the ignition (zero auxiliary heating required) locus, where inductive heating is negligible, so in this case the path taken in the plot (see Figure 5) determines the maximum heating capability required.

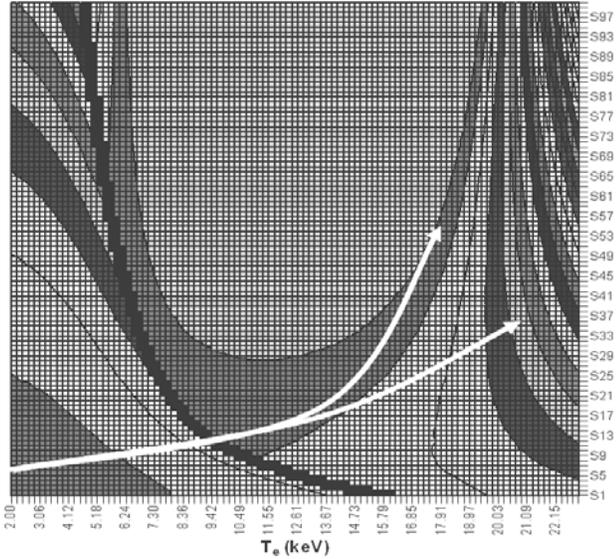

Fig. 5. A POPCON plot showing two start-up paths crossing the L-H threshold (the heavy black curve). The lower curve ends at a typical steady-state operational point, the upper curve to a near-ignition point possible in a device with no current drive. The y-axis is proportional to average plasma density, the two end-points being at 1.0 and 1.5 $\times 10^{20}/m^2$ respectively. Contours show required sustainment power, 150MW in the steady state case.

The studies undertaken suggest that this peak power is about 150 MW, occurring at the L-H transition (for which the threshold power scaling, if a meaningful concept at all, is still under development). Heating is of course easier to provide than current drive – any method can be used, provided that the power is delivered deeply enough in the plasma, whereas the need for current drive becomes a need for high current drive efficiency, experimentally found best with neutral beam injection.

Applying the 150MW intended for heating (to achieve ignition) to the flat-top burn phase for current drive was shown in this study to increase the pulse length from about three hours to about nine hours. Accordingly the number of pulses in the reactor life would be reduced from ~60,000 with no current drive (only bootstrap current) to ~20,000 with the maximum current drive efficiency presently suggested by experiments for neutral beams. This would help to reduce the design (and capital cost) problem of tolerating the fatigue caused by the large number of stress cycles associated with the pulsed machine. In 2011 when these studies were undertaken, the alternative options of ECRH and ICRH were reckoned from empirical observation to have much lower current drive efficiencies (amps/watt) and would not usefully reduce the number of pulses required in the reactor life. By the time of this work being presented at TOFE in 2012, more recent modeling had found that ECRH current drive can approach the efficiency of NBCD if high-field-side launch near the apex of the plasma is used.[8]

### VIII. FATIGUE IN THE BLANKET (KIT) AND FIRST WALL (CCFE)

A very sophisticated FEA package has been developed at KIT[9] to provide a non-linear deformation damage model, which successfully describes the behaviour of EUROFER 97 under low cycle fatigue conditions. The model uses 24 fitted parameters, each of which is in general a function of temperature and has been determined from test data on irradiated and un-irradiated material. The model reproduces the irradiation-induced reduction in fatigue lifetime at high strain, as observed in practice. At low strain ranges, the model forecasts that irradiation will allow higher fatigue lifetimes and even possibly the existence of an endurance limit, but this is not yet verified by experiments. Development of this parameterisation has facilitated a detailed fatigue analysis of the helium-cooled pebble bed ITER test blanket module, with analysis cases covering burn durations ranging from an ITER-like 400s up to 8hrs. The effects of irradiation were not included in the work so far. Significant creep-fatigue was predicted for the 8-hour pulses, while with shorter pulses the dominating factor is fatigue.

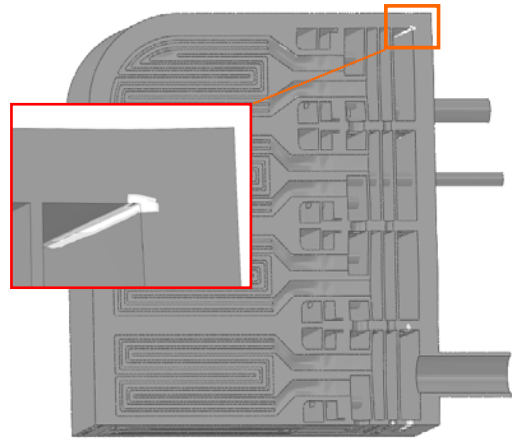

Fig. 6. Distribution of "damage" due to fatigue and creep-fatigue in helium-cooled pebble bed test blanket module after 1st pulse with 400 sec burn time. Deformations are exaggerated by a factor of 20.

A key non-linear feature of this class of analysis is that different time-histories of the neutron heat load create stresses peaking in different regions, due to the thermal wave effects, while the accumulated damage reduces the stiffness of the most stressed regions and of course creep effects are most deleterious in zones combining high stress with high temperature. Accordingly, the zones in the structure that limit the fatigue life are likely to be differently located for different circumstances, and indeed are seen to be so in this case, as shown in Figures 6 and 7.

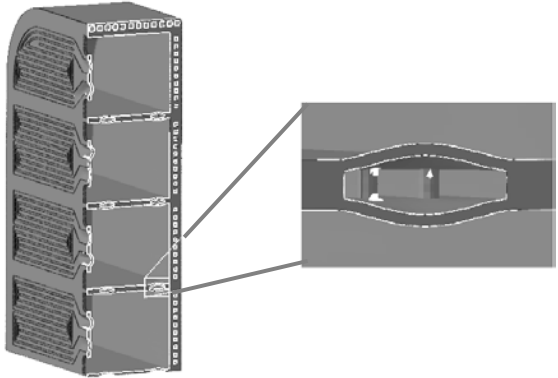

Fig. 7. As Fig.6 but now indicating the different location of the maximum creep fatigue damage region with an 8hr burn time. Deformations are exaggerated by a factor of 20.

This model forecasts that failure will only occur after ~10,000 cycles, even with 8 hour pulses, which in a reactor life of ten full-power years would require the TBM to be replaced no more than once.

The CCFE analysis adopted a different approach, analysing the first wall rather than the blanket module and using a very simple (literature-based) model of the first wall structure of parallel tubes made from EUROFER 97, taking into account fatigue but not creep fatigue.[10] The coolant flow was modelled using computational fluid dynamics in order to determine the wall heat transfer coefficients and/or the temperature distribution. The parameters in the Coffin–Manson formula modeling the S-N curves for the structural alloy of interest (EUROFER) are needed to correct for the non-zero mean stress of this load case, but are not well-determined at present.

Preliminary results indicate that with water cooling, the permissible number of cycles is more than adequate (>100,000) if the thickness of material between the coolant channel and the plasma-facing surface is 3 mm. For helium cooling, on the other hand, the permissible number of cycles is only in the low thousands due to the much greater thermal excursions and pressure variations, dominated by the pulsed $0.5MW/m^2$ surface heat load assumed throughout the CCFE calculations.

It is clear from these two studies that one of the original aims, the generation of simple formulae implementable in system codes for estimating the lifetime of the first wall and blanket, is a non-trivial task.

## IX. IMPROVED $Nb_3Sn$ CONDUCTOR DESIGNS (CRPP & ENEA)

$Nb_3Sn$ is a brittle inter-metallic compound, so stresses imposed after creating it via the necessary high temperature reaction must be minimised. During the reaction the material anneals so all strains disappear, but upon cooling from heat treatment temperature (~920K) to cryogenic operating temperature (~4.2K), compressive strains of up to ~0.7% are created by the greater thermal contraction of the surrounding material. This can include the contraction of the conduit if a cable-in-conduit conductor (CICC) was created before reacting. (The tensile stress created by use of the conductor to produce a high magnet field can offset this, although only by ~0.1%, since within a limited range the effect of these elastic strains on critical current is reversible.) In the React and Wind method considered by CRPP, the conductor can be cooled to room temperature without the presence of a constraining jacket, allowing most of the cool-down strain to be eliminated. This results in a 30-40% better final performance, provided the conductor can be unwound from the spool and wound into the final shape without incurring irreversible bending strain.

In comparison with the alternative Wind & React method, the size of the reaction furnace when using React and Wind is smaller. Moreover, the mass of material to be heat treated is an order of magnitude smaller (in the case of a TF coil) because there is no conduit during heat-treatment and no strong mould necessary to maintain the D-shape of the winding (within tight tolerances) during warm-up and cool-down.

The electric field criterion ($E_{crit}$) employed to define the reversibility range has evolved over the decades, from 100µV/m in the 1980s, towards only 1-2 µV/m for large coils today. This has led to more careful determination of the critical current $j_{crit}$ and the n-index (a measure of the sharpness of the superconducting transition, i.e. $E/E_{crit} = (j/j_{crit})^n$ ) of CIC conductors. Comparison with a free-standing strand then reveals the extent to which degradation occurs in the cable during manufacturing, winding, cooling down, initial training and subsequent cycling of the magnet.

Most conductor testing to date extends only to 1000 cycles. The test results of the ITER central solenoid (CS) conductor samples in 2010-2011 show that the performance degradation upon cyclic load does not asymptote after a few hundred cycles, as previously interpreted. Measurements on the JACS2 sample, which was made from a high performance strand, show that the degradation rate is small if considered over an interval of 100 cycles, but continues (without saturation) over 15000 cycles. Several variations in the cable design have been tried to mitigate the degradation, including reduction of void fraction and use of long twist pitches, but the lifetime of the 2010-2011 version of the ITER CICC remained limited to a few thousand full load cycles even in the best case, and is therefore not considered suitable for a pulsed version of DEMO. Of course work to identify better optimised strand and cable configurations for ITER continues, e.g. considering very short as well as very long twist pitch, and increasing the ratio of superconductor to copper in the cable.

The cause of irreversible degradation was studied in 2002 by testing individual strands extracted from the CICC tested in SULTAN (CRPP-Switzerland). The strands extracted from highly loaded CICC sections showed much poorer performance than strands extracted from low load sections. Microscopic investigations confirmed that strand cracking and filament breakage occurs, as shown in Figure 8.

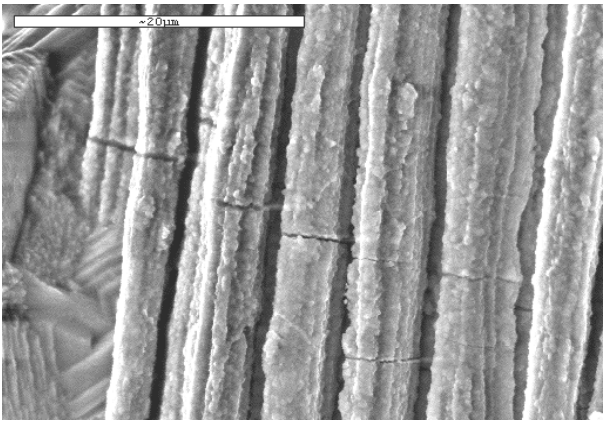

Fig. 8. Scanning microscope image of etched filaments showing crack propagation from the left (tensile region).

The location of the breakages shows that bending, with the associated tension on one side of the bent filament, is the primary reason for irreversible degradation. The need to keep every strand close to the neutral axis of the bending incurred during winding limits the thickness of the cable, leading towards flat ribbon cable designs which distribute the electromagnetic load over a broad surface, reducing the accumulated pressure on the strand cross-overs. A 45kA flat cable in conduit conductor was developed and tested at CRPP in 2005-2006.[11]

In some design variants, the cross-section of the helium cooling channel is far larger than in ITER conductors, but the helium surrounding each strand is nearly stagnant, so analyses of the mechanical stability of a self-supported coil, and of the cooling performance of a low-void-fraction cable are required. Additional copper for quench protection can be provided in the form of wires wrapped around the superconducting ribbon cable, since bending stresses in the copper do not alter its electrical behaviour. An example of such a rectangular conductor concept is sketched in Figure 9.

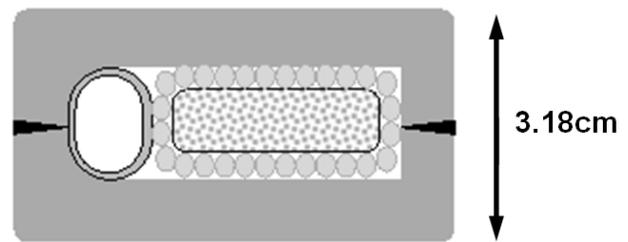

Fig. 9. Example design of a flat ribbon cable soldered into a thin sleeve, surrounded by copper wires (the grey circles) for improved quench protection. The helium cooling pipe is at left and the jacket is stainless steel.

ENEA concentrated on the Wind & React CICC system, as used at present for ITER. This makes it possible to avoid the large post-reaction thermal strain, but only if the conduit has a low coefficient of thermal expansion. When the role of the conduit is mainly hydraulic containment as in the ITER TF conductor, the ideal material is titanium. If the conduit material must also take the operating loads as in the ITER CS, the ideal material is Incoloy or Invar. The technology for manufacturing coils based on Ti and Incoloy jacketed conductors has been developed and fully demonstrated in prototypes considered for ITER.

ENEA analysis indicates that given a rectangular cable with no cooling channel, TF coils like those of ITER could be built without the radial plates found necessary in the ITER design. Figure 10 shows the ENEA rectangular conductor and the results of trials showing no degradation of current-sharing temperature (near-quench conditions) when the orientation of the conductor is correct.[12]

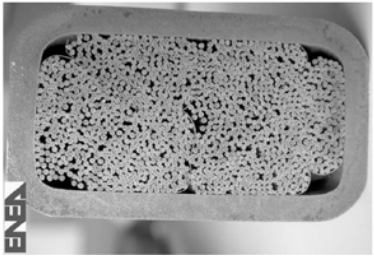

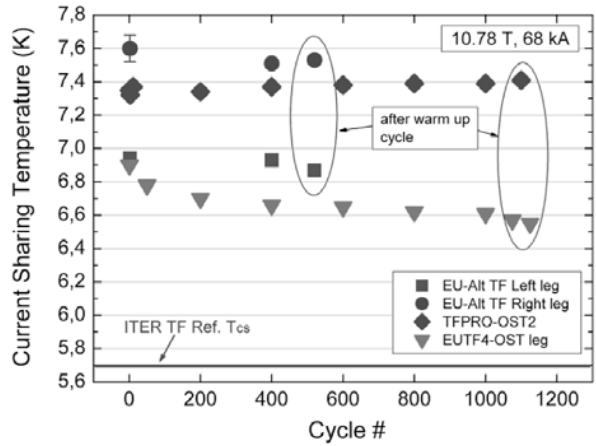

Fig. 10. Upper: photograph of the cross-section of an ENEA rectangular conductor;

Lower: the effect of conductor orientation on current sharing temperature when operationally cycled. The upper points (circles and diamonds) have the Lorentz force parallel to the short axis of the conductor; the lower points (squares and triangles) have it parallel to the long axis.

The following list represents a compilation of the recommendations by the two groups for $Nb_3Sn$ CICC fatigue life improvements, based on the test results and understanding pertaining at the beginning of 2012:

- To remove the heat generated by nuclear radiation, AC losses, joint resistance, etc., forced flow of supercritical helium is needed.
- The void fraction must be minimised so that each strand is supported as fully as possible.
- Each magnet should include different grades of conductor (high field, low current density / low field, high current density).
- Longer cable twist pitches should be introduced (although this is expected to make AC losses worse),
- A rectangular cable cross-section should be adopted.

Some observers have a more positive view of the prospects for the development of high temperature superconductors (HTS) than the acknowledged experts in fusion magnet design, since their development is currently meteoric due to very large industrial investment for non-fusion applications. However inevitably those other applications do not entail measurable neutron fluence, nuclear heating, or significant electromagnetic load variations and so further testing and development is likely to be necessary for fusion purposes once otherwise suitable HTS conductors have been developed.

## X. CONCLUSIONS

Interest in pulsed versions of a DEMO tokamak reactor is being driven by concerns about the large recirculating power necessary for the plasma current drive system of a steady-state version, which drives up the total fusion power requirement and hence the divertor heat loads unless relatively advanced divertor configurations or large radiative power fractions are invoked. Alternatively a high bootstrap fraction could be used as a design premise but this in turn necessitates carefully tailored pressure and current profiles, perhaps difficult to sustain in a device which would then be dominated by alpha heating with its risk of self-organisation. Thus while steady state remains the ultimate goal, a pulsed tokamak reactor might offer "net electricity into the grid" with less challenging technology and physics (detailed plasma control) requirements.

The fusion power of DEMO depends sensitively on whether or not steady state operation is required. A steady-state machine needs ~3 GW or more of fusion power, to support the plasma current drive system and balance of plant with a reasonable overall recirculating power fraction. A pulsed machine with 3GW of fusion power would achieve a net electrical output greater by perhaps >200MW (the power not needed for current drive). An operational option was shown to be running the auxiliary heating system needed for start-up to provide some current drive, extending the pulse length and hence reducing the number of fatigue cycles in the life of the reactor by a factor of about three. Alternatively, variants of a pulsed machine could demonstrate generation of electrical power at only 1 GW of fusion power by running at high Q with no current drive, consistent with increased margins for the superconductors and lower divertor loading, albeit with a short pulse length ~2 h.

Disruptions that are not effectively mitigated will cause surface melting even in tungsten. Wall erosion by fast neutrals and divertor erosion by ions during routine operation are also significant.

It is not clear whether energy storage on-site is mandatory for a pulsed reactor, but it is likely that storage will be required to cushion the grid from the rapid ramping at the beginning and end of each pulse, and to provide start-up power.

A pulsed reactor with no current drive can be operated without large additional heating or current drive power during the burn phase, but heating is required for start-up. Preliminary work indicates that the peak power requirement is about 150 MW, necessary at the L-H transition. This is only a little less than the 200 MW of current drive power representing the lower end of the range required for steady state tokamak reactors, but heating is easier to provide than current drive and the pulsed machine could be operated at very high Q, greatly reducing the heating power required during the burn.

Models of the fatigue of the blanket and first wall have been made, but are not yet able to give definitive results. Fatigue in the CS and TF coils and their support structures has not yet been studied in detail, but it is seems likely that high performance materials will need to be used. Future work should include the effects of irradiation and creep fatigue in the first wall and blanket, and should ideally generate simple formulae, suitable for system codes, for estimating the fatigue lifetime of all the cyclically stressed components.

The $Nb_3Sn$ conductor-in-conduit cable intended to be used for ITER (in its 2011 version) seems unlikely to be suitable for a pulsed DEMO. Improved conductors may be of the React & Wind or Wind & React approach, but would benefit from lower conductor void fraction, markedly different cable twist pitches, and a rectangular cross-section. The options of incorporating radial plates like those in the ITER TF coils and of having the helium cooling channel within the conductor may well not be appropriate for DEMO and beyond.

It would appear that currently the most technically challenging of the pulsed reactor issues studied for this work are disruptivity and the consequences of any major disruptions, and creep fatigue in the irradiated materials of the first wall and blanket. The other design aspects considered appear to be tractable to design optimizations, or affect the cost and market acceptability of the reactor – both important factors for the eventual realization of fusion power of course - rather than the technical feasibility.

## ACKNOWLEDGMENTS

The EFDA PPPT work reported here was carried out by the following:

CEA     J.L. Duchateau & P. Hertout
CRPP    P. Bruzzone
CCFE    R. Clarke, W. Han, P. Knight, T. N. Todd & Z Vizvary
ENEA    L. Muzzi, F. Orsitto & G. M. Polli
KIT      J. Aktaa & Y. Igitkhanov

The CCFE aspects of this work were funded by the RCUK Energy Programme under grant EP/I501045 and the European Communities under the contract of Association between EURATOM and CCFE. The views and opinions expressed herein do not necessarily reflect those of the European Commission. This work was carried out within the framework of the European Fusion Development Agreement.